# Imaging the potential distribution of individual charged impurities on graphene by low-energy electron holography

Tatiana Latychevskaia*, Flavio Wicki, Conrad Escher, Hans-Werner Fink

Physics Department, University of Zurich,

Winterthurerstrasse 190, 8057 Zurich, Switzerland

*Corresponding author: tatiana@physik.uzh.ch

## ABSTRACT

While imaging individual atoms can routinely be achieved in high resolution transmission electron microscopy, visualizing the potential distribution of individually charged adsorbates leading to a phase shift of the probing electron wave is still a challenging task. Low-energy electrons (30 – 250 eV) are sensitive to localized potential gradients. We employed low-energy electron holography to acquire in-line holograms of individual charged impurities on free-standing graphene. By applying an iterative phase retrieval reconstruction routine we recover the potential distribution of the localized charged impurities present on free-standing graphene.

## 1. INTRODUCTION

The charge density, respectively the electrostatic potential of atoms in crystals can directly be probed by x-ray [1-2] or electron [3-5] scattering. However, such diffraction experiments are limited to crystals, and the challenging task is to visualize the charge density or electrostatic potential of an individual adsorbate. Recently, by combining techniques such as high-resolution transmission electron microscopy with first-principles electronic structure calculations [6] or scanning tunneling microscopy with Raman spectroscopy, or x-ray spectroscopy with first principles calculations [7] it has been demonstrated that charge redistribution due to chemical bonding of nitrogen dopants to graphene could be revealed.

Adsorbates change the electronic properties of graphene thus making graphene a promising candidate for highly-sensitive electrical gas sensors. The conductivity $\sigma$ of graphene can be expressed as $\sigma = n\mu e$ where $n$ is the carrier density, $\mu$ is the mobility and $e$ is the electron charge, respectively. Thus, a change in conductivity can be either caused by a change in the carrier density or in the mobility of the charge carriers.

Density Functional Theory (DFT) calculations show that most adatoms transfer electrons to graphene [8], initiating a charge redistribution in their vicinity. When charge is transferred from the adsorbate to graphene, the adsorbate constitutes a charged impurity. The charge density redistribution between the adsorbate and graphene can be complex and exhibiting alternating regions of electron density depletion and enrichment [9]. For example, DFT simulations for a $CO_2$ molecule on graphene demonstrate a strong localization of electron density depletion, that is, a positively charged impurity which extends over 2.7 nm in radius [9]. Such charged impurity creates a Coulomb potential at the position of the impurity which is strong enough to block carrier diffusion. Hwang *et al* provided a theoretical study and showed the inverse dependency of the mobility $\mu$ on the density of impurities $\mu = 1/n_{\text{impurities}}$ [10]. Chen *et al* measured the increase in the mobility $\mu$ when the concentration of dopant (potassium) increases [11]. However, no decrease of the mobility has been observed in other experiments for small molecules like NO, $NO_2$, and $NH_3$ [12-16], different models were proposed to explain this discrepancy [17]. The latest results show that charge carrier mobility remains almost constant for non-covalent functionalization of graphene, and it is reduced for covalent functionalization [18-19]. It has recently been discussed that charged impurities are the predominant source of disorder in most graphene samples leading to a limited charge carrier mobility ([20] and references therein), vital for graphene applications in mesoscopic devices.

So far, the detection of individual adsorption events was demonstrated by two groups. Schedin *et al* in 2007 reported individual absorbing events of nitrogen dioxine molecules on graphene by detecting a change in carrier density in graphene measured as changes in the Hall resistivity [16]. Sun *et al* in 2016 reported individual events of carbon dioxide molecules physisorbed on double layer suspended graphene by detecting a step-like change in the resistance [9]. The detection of individual adsorption events remains the goal of high-resolution gas sensing.

Theoretical simulations for electron scattering and absorption by an individual atom on graphene are complicated if not impossible to perform. The phase shifts that form the scattering amplitude expansion are conventionally obtained for a certain model of the scattering potential [21]. The later in turn should be built by taking into account the electron distribution of the atom. The task becomes particularly difficult when the atom is charged, leading to re-distribution of electrons within the atom. Here, we would like to cite John Pendry: "The ion-core potential acts on conducting electrons influencing their motion, and therefore changes their charge density. The problem is a self-consistent one and indeed when accurate band structure calculations are needed an iterative procedure for calculating ion-core scattering conduction electron density, and the ion core scattering again with the new screening charge, must be employed: a tedious procedure" [22]. And even if such potential of a charged atom was available, it would be again only a rough approximation. A correct model must take the graphene support into account with all its imperfections such as adsorbates, vacancies etc. The potential distribution associated with an individual charged atom appears therefore impossible to accurately predict by theory. However, when making a number of assumptions, as for example the exact orientation of an adsorbate on graphene, it is possible to perform DFT simulations for an individual adsorbate and to provide the charge density redistribution [6, 9]. We demonstrate experimental observation of an individual charged impurity by means of low-energy holography and recover its projected potential from its holograms by an iterative phase retrieval reconstruction.

## 2. IMAGING WITH LOW-ENERGY ELECTRONS

Recently we reported that low-energy electrons (30 – 250 eV) exhibit high sensitivity to local electric fields and thus can be employed to directly visualize individual charged adsorbates [23]. The charged adsorbates were identified by their appearance in the hologram as dark (for a negatively charged adsorbate) or as bright (for a positively charged adsorbate) spots. It was also demonstrated that a conventional reconstruction routine fails to retrieve the distribution of the charged adsorbates [23]. In this study we show that an iterative procedure allows reconstruction of the absorption and phase distribution of the adsorbate. The potential distribution of an adsorbate can then be extracted from the retrieved adsorbate's phase distribution.

In order to understand what constitutes the absorption and phase contrast in the observed patterns, we briefly address the processes involved when a low-energy electron is interacting with the sample [22]. Firstly, an incident electron can be elastically scattered by a localized potential within the sample. In fact, the scattering for low-energy electrons exhibits pronounced maxima of the scattering amplitude in forward and backward directions [21]. Moreover, electron energy losses due to excitations of phonons or plasmons are conceivable as well as absorption of electrons. The final distribution of the object wave respectively its signature at the detector is the time-average over these two main processes, forward scattering and absorption. Inelastic scattered electrons do not contribute to the interference pattern but rather to the almost uniform background. The interaction between an electron beam and a charged object has been previously discussed in the literature, for example in connection with the Möllenstedt biprism effect [24]. A charged object deflects even those electrons which are passing the object at some distance. Thus the phase of the electron wave is altered. This change in the phase of the electron wave can be characterized by a phase-changing distribution superimposed onto the wave in the object plane. For example, in the case of a charged wire, the introduced phase change distribution is analogous to a phase change that a light wave experiences while passing through an optical prism, hence the name electron "biprism effect".

## 3. EXPERIMENTAL SETUP

A typical setup for low-energy electron holography [25] is depicted in Fig. 1 and explained in detail in [23, 26]. Electrons are field emitted from a sharp tungsten tip [27-28]. The electron wave passes through the sample while part of the wave is scattered by the object. Interference between the scattered and non-scattered wave leads to a hologram recorded at a distant detector. The sample is free-standing graphene with some residual adsorbates after sample preparation. The graphene samples were prepared as described elsewhere [29]. Since transparent free-standing graphene provides an overall equipotential plane, a neutral adsorbate present on graphene results in an ordinary hologram, as illustrated in Fig. 1(b). However, a charged impurity locally creates a high electric field deflecting the passing electrons. The presence of a positively charged impurity thus leads to a distinctive signature in the hologram, a bright spot [23], as illustrated in Fig. 1(c).

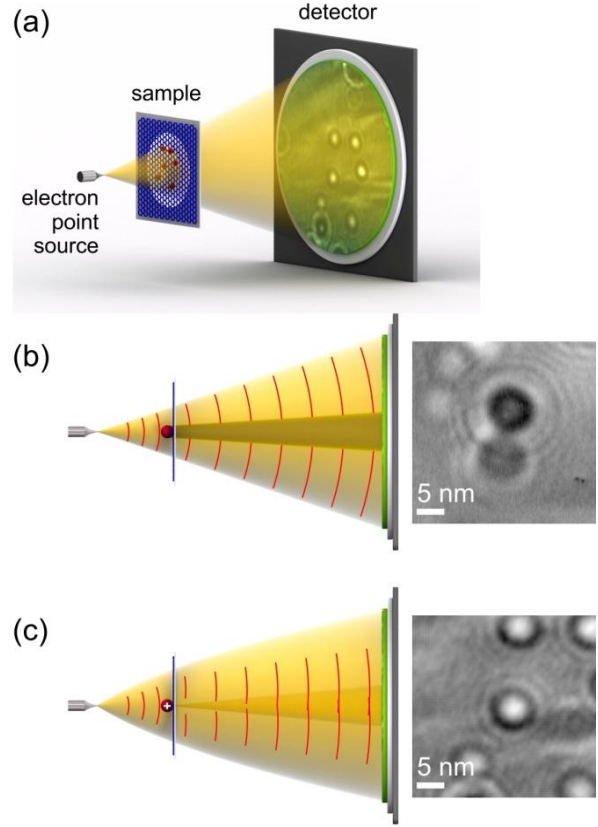

FIG. 1. Experimental arrangement. (a) Low-energy electron holographic microscope. The distance from source to sample is in the range of tens of nanometres and the distance from source to detector amounts to 47 mm. (b) Imaging of a neutral adsorbate whereby part of the beam is absorbed resulting in an ordinary hologram. (c) Imaging of a positively charged impurity. The electron trajectories are deflected inward by the presence of a positive charge resulting in a bright spot on the detector.

## 4. RECONSTRUCTION OF HOLOGRAMS

The transmission function in the object plane $(x, y)$ can be written as

$$t(x, y) = \exp[-a(x, y)] \exp[i\varphi(x, y)], \tag{1}$$

where $a(x,y)$ describes the absorption and $\varphi(x,y)$ the phase distributions respectively. Absorption $a(x,y)$ is related to the loss of elastically scattered amplitude, it does not involve any inelastic process. The transmission function can always be represented as

$$t(x,y) = 1 + o(x,y), \qquad (2)$$

where 1 represents the part of the exit wave that is the same as the incident wave, and $o(x,y)$ represents the part of the exit wave that has been modified by the presence of the object. For example, in the particular case of a weak-phase object, it can be written as $t(x,y) \approx 1 + i\varphi(x,y)$ and thus $o(x,y) = i\varphi(x,y)$. However, we must note that in the case considered here, no such approximation is applied, so that

$$o(x,y) = \exp[-a(x,y)]\exp[i\varphi(x,y)] - 1. \qquad (3)$$

The mathematical separation of the transmission function into two terms (Eq. (2)) helps to separate the contribution of the reference from the one of the object wave. In the detector plane, the wave described by the first term in Eq. (2) provides the reference wave $R(X,Y)$, and the wave described by the second term in Eq. (2) provides the object wave $O(X,Y)$.

In the case of a non-charged object, the object distribution $o(x,y)$ is finite in physical size and smaller when compared to the extent of the reference wave. Thus, the fulfilment of the condition $|O(X,Y)| \ll |R(X,Y)|$ together with a well-defined reference wave ensure that such a hologram can be reconstructed [30].

In the case of imaging a charged object with an electron wave, the distribution of the electron wave is altered by the object. A charged impurity deflects even those electrons which are passing the localized charge at some distance. Therefore, a clean separation between the reference wave (not affected by the object) and the object wave (affected by the object) is not possible any more. Consequently, the lack of a well-defined reference wave precludes an ordinary hologram reconstruction. A sensible hologram reconstruction becomes thus more involved. The problem of object reconstruction from the recorded interference pattern is similar to the phase retrieval problem in coherent diffraction imaging (CDI), where only the intensity of the scattered wave is detected but no reference wave is present [31]. In CDI, typically, an iterative procedure is applied to retrieve the missing phase distribution in the detector plane and consequently to reconstruct the object distribution. We therefore apply such iterative procedure to reconstruct holograms of charged impurities. The iterative reconstruction procedure applied here is based on the procedure described elsewhere [32-33], starting with an initial random phase distribution in the detector plane; details of the iterative reconstruction procedure are presented in Appendix 1.

An electron hologram where several bright spots are observed was selected. To begin with, the conventional non-iterative reconstruction procedure was carried out for different source-to-sample distances ranging from 10 nm to 1000 nm. At a source-to-sample distance of 82 nm the reconstruction of objects such as clusters was found to be compliant; hence, this distance was identified as the correct source-to-sample distance. In the reconstruction obtained in such a way, the bright spots did not converge to a meaningful object reconstruction. Four selected bright spots are shown in Fig. 2(a). The spots are selected as to all being located at a relatively close distance to each other varying from 11.6 to 20.6 nm. The proximity to each other ensures that these impurities are imaged under similar conditions. Each bright spot was reconstructed separately by applying an iterative procedure as explained in Appendix 1. For each spot, 100 iterative runs were performed with an initial phase that was randomly distributed. The outcome of each iterative run was essentially identical. The error of the fitted intensity was evaluated for each iterative outcome as following:

$$Er = \frac{1}{N^2} \sum_{i,j=1}^{N} \left| \frac{I_{\text{measured}}(i,j) - I_{\text{iterated}}(i,j)}{I_{\text{measured}}(i,j)} \right|, \qquad (4)$$

which is an estimate of the difference between the measured and the iterated values in relation to the measured values. The mean errors estimated for 100 iterative runs for each spot are small: $\langle Er_1 \rangle = 7.074 \times 10^{-2}\%$, $\langle Er_2 \rangle = 9.626 \times 10^{-2}\%$, $\langle Er_3 \rangle = 9.094 \times 10^{-2}\%$ and $\langle Er_4 \rangle = 7.998 \times 10^{-2}\%$, where $\langle ... \rangle$ denotes averaging over 100 iterative runs. The corresponding standard deviations of the errors for the four spots are negligibly small: $\sigma_{Er,1} = 1.010 \times 10^{-6}\%$, $\sigma_{Er,2} = 1.014 \times 10^{-7}\%$, $\sigma_{Er,3} = 1.720 \times 10^{-9}\%$ and $\sigma_{Er,4} = 1.305 \times 10^{-9}\%$. To cross-check our results we also used the same iterative phase reconstruction but with an initial phase distribution that follows a phase distribution caused by a Coulomb potential, with added noise. The resulting reconstruction was identical to the one obtained with a completely random initial phase distribution. Such good reproducibility can be explained by the fact that in in-line holography the size of the imaged object is typically much smaller than the area covered by the reference wave. Thus, the number of pixels that sample the measured intensity (the number of equations) exceeds the number of pixels that are required to sample the unknown object absorption and phase distribution (the number of unknowns). This condition also holds when the object distribution has an extended phase-changing distribution, as for example in the case of an electron wave affected by a localized potential of a charged object [33], more details are provided in Appendix 1.

The iteratively fitted intensity distributions are shown in Fig. 2(b). The radial profiles of the measured and the iteratively fitted intensity distributions are in good congruence as shown in Fig. 2(c).

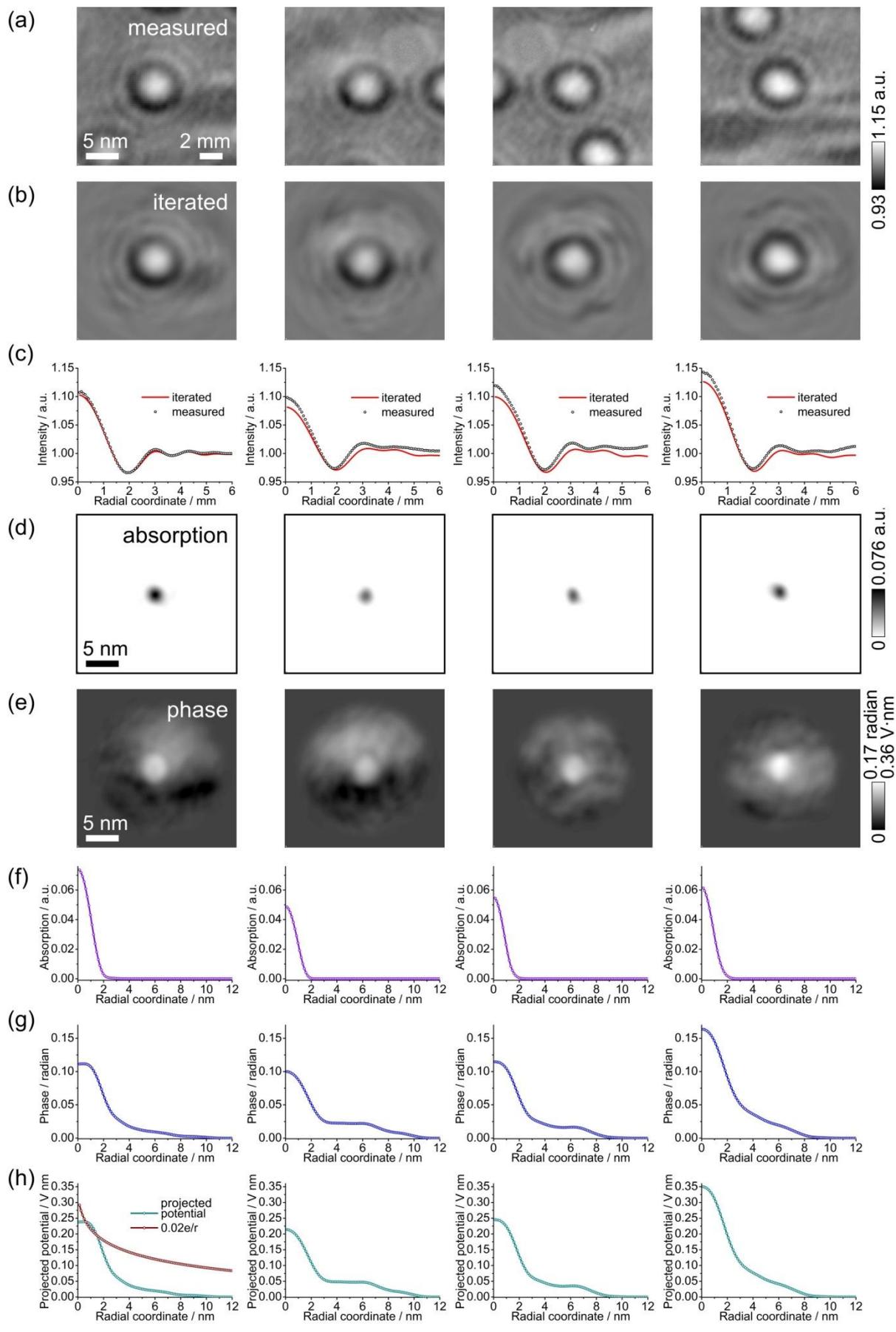

FIG. 2. Iteratively reconstructed absorption and phase distribution of individual charged impurities. (a) Four selected bright spots in a hologram acquired with electrons of 30 eV kinetic energy. (b) The related intensity distributions obtained after 2000 iterations. (c) The radial profiles of the measured and iteratively obtained intensity distributions. (d) and (e) The related iteratively reconstructed absorption and phase distributions, (f) and (g) their radial profiles. (h) The radial distributions of the projected potentials obtained from the reconstructed phase distributions. In the first inset, a projected potential of a 0.02$e$ charge is shown for comparison.

## 5. DISCUSSION

Figure 2 shows reconstructed absorption and phase distributions. The absorption distributions were fitted with a Gaussian function, the following standard deviations were obtained: $\sigma_1 = 8.78$ Å, $\sigma_2 = 6.99$ Å, $\sigma_3 = 6.31$ Å, and $\sigma_4 = 7.80$ Å. These values are relatively large when compared to the empiric covalent radii of typical adsorbates found on graphene: 70 pm for carbon, 25 pm for hydrogen, 60 pm for oxygen and 110 pm for silicon [34]. The discrepancy can be explained by the limited resolution of the low-energy electron holographic setup. The intrinsic resolution of the setup is described by the formula:

$$R_{\text{intrinsic}} = \frac{\lambda}{2\text{NA}}, \qquad (5)$$

where NA is the numerical aperture of the setup. The intrinsic resolution of the setup with a detector diameter of 75 mm at the wavelength of 0.22 nm for 30 eV electrons amounts to $R_{\text{intrinsic}} = \frac{\lambda}{2\text{NA}} = 0.23\,\text{nm}$. However, the resolution is degraded by residual mechanical vibrations smearing the fine interference fringes in the hologram thus leading to the decrease of the effective size of the hologram, which in turn leads to a blurry appearance of the reconstructions. Moreover, the fact that the reference wave does not coherently illuminate the entire detector, also contributes to a limited experimental resolution. The maxima of the retrieved absorption distributions are relatively small: $a_1 = 0.076$, $a_2 = 0.053$, $a_3 = 0.057$ and $a_4 = 0.067$. When the adsorbate is not charged, the effect of absorption alone results in a hologram which exhibits a very weak contrast not distinguishable from the noise level.

As evident from the reconstructions shown in Fig. 2, the reconstructed absorption distributions (Fig. 2(d)) appear to be narrower than the reconstructed phase distributions (Fig. 2(e)). This agrees well with the notion that the phase distribution unlike the absorption distribution does not reflect actual size of the adsorbate itself but rather its potential distribution caused by the charge.

The reconstructed phase shift distribution $\Delta\varphi(x, y)$ can be interpreted as a projected potential [35], see Appendix 2:

$$\Delta\varphi(x, y) = \frac{2\pi e}{h\upsilon} V_{\text{proj}}(x, y), \tag{6}$$

where $e$ is the elementary charge, $h$ is the Planck's constant, $\upsilon = \sqrt{2eU/m_e}$ is the speed of the probing electrons, $U$ is the accelerating voltage, $m_e$ is the electron mass and $V_{\text{proj}}(x, y)$ is denoting the projected electrostatic potential:

$$V_{\text{proj}}(x, y) = \int_{\text{path}} V(x, y, z)\, dz, \tag{7}$$

where $V(x, y, z)$ is the electrostatic potential along the path of the probing electrons.

A phase distribution reconstructed from an interference pattern is unambiguous except for any constant offset. We selected the offset in such a way that the minimum of the phase distribution, and therefore of the projected potential, is zero. Thus, it is impossible to uniquely retrieve the constant value which the potential reaches at a large distance from the charge.

This constant value however should correspond to the thermal energy, which amounts to 25.7 meV at room temperature.

The distributions of the projected potential were obtained from the phase distributions using Eq. (6) and are shown in Fig. 2(h). It is evident that the potential of a charged adsorbate is almost constant within the area occupied by the adsorbate, reaching about $(V_{proj})_{max} = 0.25 - 0.35$ V·nm. The potential slowly decays outwards towards 0 within a 9 – 10 nm distance from the adsorbate. Similar behaviour of the potential was reported for $CO_2$ molecule on graphene whose simulated potential reached a constant value of the thermal energy at about 2.7 nm distance from the molecule [9]. For comparison, we selected a Coulomb potential of the form $V(r) = 0.02e/r$ and simulates the corresponding projected potential, shown in Fig. 2(h) (first panel on the left). The corresponding simulated projected potential exhibits a faster decay when compared to the experimentally obtained projected potential. Thus, the realistic potential distribution associated with a complex charge density redistribution between the adsorbate and graphene [9] can only roughly be approximated by a $1/r$ dependency [23] that corresponds to a free charge and disregards any shielding due to an electronic rearrangement within the graphene. The simulated projected Coulomb potential does not reach zero as the experimentally measured projected potential does. This is simply due to having set the constant value which the potential reaches at large distance from the charge to zero in the experimental case.

## CONCLUSIONS

We demonstrated that projected potentials of individual adsorbates can be obtained from their low-energy electron holograms by iterative phase retrieval reconstruction. A stable solution can be obtained provided the size of the adsorbate is much smaller than the size of the imaged area. The retrieved projected potential of an individual adsorbate exhibits a plateau in the region of the adsorbate's location where the potential reaches the maximum of about 0.25 – 0.35 V·nm. When compared to a simulated projected Coulomb potential, the projected potential recovered from an experimental measurement displays a much slower decay, turning into a constant value at about 10 nm from the impurity position. Low-energy electron holography thus offers a unique possibility to quantitatively evaluate the potential distribution around an individual adsorbate. Provided that the chemical identity of the adsorbate can be controlled, further experiments to characterize the potential distribution of specific molecules become possible.

**APPENIX 1**

A stable solution can be obtained by an iterative approach when the number of unknowns equals the number of equations. In CDI this condition is achieved by sampling the diffraction pattern at a rate corresponding to at least twice the Nyquist frequency, which in the object plane turns into a known support that exceeds twice the object size in each dimension [36]. In in-line holography, the condition of sampling is less strict, because of the well-defined reference wave. Though when imaging a phase-changing object, the number of knowns is reduced. However, it has previously been demonstrated that phase-changing objects can be retrieved from their in-line holograms by phase retrieval methods [33, 37]. This fortunate situation can be explained in the following way. An in-line hologram is sampled with $N \times N$ pixels, thus there are $N \times N$ measured intensity values providing $N^2$ equations. The number of unknowns is given by the number of pixels that are required to sample the unknown absorption and phase distributions in the object plane. Since the object occupies a finite region, its absorption distribution is only non-zero within a $N_0 \times N_0$ pixels region. This leaves the number of unknowns equal to $N^2 - N_0^2$. The unknown phase-changing distribution thus can be sampled with $N^2 - N_0^2$ pixels corresponding to the remaining number of the unknowns.

In the case when imaging an atom or a small molecule within a total field of view (FOV) of the order of a few tens of nanometers, the object occupies an area which is much smaller than the FOV. The phase-changing distribution in the object plane is provided by the Coulomb potential of an individual impurity. Therefore, the phase-changing distribution in the object plane also has a finite size, although its distribution extends over a much larger area than the physical size of the absorber. For example, for the parameters in this study, the field of view is $48.8 \times 48.8$ nm$^2$, the area occupied by a 1 nm radius absorbing object amounts to 3.14 nm$^2$ which translates into a ratio of unknowns/knowns=0.0013. The ratio of the unknown phase distribution to the total area is thus 0.9987. For an object with a size much smaller than the imaged area, the unknown phase distribution can be almost as large as the total imaged area. In reality, the extent of the meaningful phase-changing distribution is typically smaller than the FOV, and thus the number of unknowns is reduced. This ensures a quick convergence of an iterative reconstruction routine to the best fitting solution, as exemplified for example in the reconstruction of a charged sphere from its in-line hologram [33].

The iterative reconstruction procedure applied here is based on the procedure described elsewhere [32-33]. The reconstruction starts in the hologram plane. In the case of a non-charged object, the phase of the reference wave is constant in the detector plane and the iterative reconstruction can begin with the initial phase distribution in the detector plane set to zero [32]. In the case of a charged object, the distribution of the reference electron wave is altered by the object, and therefore the phase of the reference wave and the phase of the total wavefront in the detector plane are both unknown. Therefore, for the current study we set the initial phase distribution on the detector plan to be randomly distributed in the range $\left[-\frac{\pi}{2}, \frac{\pi}{2}\right]$.

(i) The hologram is reconstructed and the transmission function in the object plane $t(x,y)$ is extracted using an algorithm for hologram reconstruction described elsewhere [38-39]. The transmission function in the object plane is given by $t(x,y) = \exp[-a(x,y)]\exp[i\varphi(x,y)]$, where $a(x,y)$ denotes the absorption and $\varphi(x,y)$ the phase distributions respectively.

(ii) In the object plane the following constraint is applied: $a(x,y) \geq 0$ based on a physical notion that the absorption cannot be negative [32]. Next, negative values of $a(x,y)$ are thus replaced with zeros, and an updated $a'(x,y)$ is obtained in this way. No constraint is imposed on the phase distribution $\varphi(x,y)$.

(iii) Next, the wavefront distribution in the detector plane is simulated with the updated transmission function $t(x,y) = \exp[-a'(x,y)]\exp[i\varphi(x,y)]$ [39].

(iv) The amplitude of the simulated wavefront is replaced with the measured amplitude given by the square root of the measured hologram intensity. The phase distribution of the simulated wavefront is adapted for the next iteration starting at (i).

2000 iterations were applied in total. The recovered absorption distribution was smoothed to avoid accumulation of noisy peaks. During the first 1000 iterations, the absorption distribution starts to form a pronounced peak with a width of about 2 nm. For the remaining 1000 iterations a loose support of 5 nm in radius was applied to suppress artefacts around the absorption reconstruction.

## APPENIX 2

A plane wave impinging onto an object with the electric potential $V$ and/or a magnetic potential $\vec{A}$ experiences a phase modulation. To obtain the phase of the exit wave, the so-called eikonal equation along all possible trajectories is evaluated:

$$\varphi_{obj} = 2\pi \int_{path} \vec{k}_{obj} \, d\vec{s}, \tag{8}$$

where

$$\vec{k}_{obj} = \frac{\sqrt{2em(U+V)}}{h} \vec{e}_p - \frac{e}{h} \vec{A}, \tag{9}$$

with $\vec{e}_p$ being the unit vector along the trajectory. In the absence of the object, we obtain:

$$\varphi_{vac} = 2\pi \int_{vac} \vec{k}_0 \, d\vec{s}, \tag{10}$$

where

$$\vec{k}_0 = \frac{\sqrt{2em_e U}}{h} \vec{e}_p. \tag{11}$$

Here $e$ is the elementary charge, $h$ is the Planck's constant, $\upsilon = \sqrt{2eU/m_e}$ is the speed of the probing electrons, $U$ is the accelerating voltage and $m_e$ is the electron mass.

The phase shift due to the electrostatic potential $V$ when comparing it to the vacuum is thus given by

$$\begin{aligned}\Delta\varphi_{obj} &= 2\pi \int_{path} \left( \frac{\sqrt{2em_e(U+V)}}{h} - \frac{\sqrt{2em_e U}}{h} \right) ds = \\ &= 2\pi \frac{\sqrt{2em_e U}}{h} \int_{path} \left( \sqrt{\frac{U+V}{U}} - 1 \right) ds \approx 2\pi \frac{e}{h\upsilon} \int_{path} V \, ds,\end{aligned} \tag{12}$$

where we applied the approximation $U \gg V$ and introduced $\upsilon = \sqrt{2eU/m_e}$ as the speed of the probing electrons.